\documentclass[usenatbib]{mn2e}
\usepackage{graphicx,epsfig,natbib,amssymb}
\usepackage{wrapfig}
\def\gtwid{\mathrel{\raise.3ex\hbox{$>$\kern-.75em\lower1ex\hbox{$\sim$}}}}
\def\ltwid{\mathrel{\raise.3ex\hbox{$<$\kern-.75em\lower1ex\hbox{$\sim$}}}}
\def\lessim{\mathrel{\raise.3ex\hbox{$<$\kern-.75em\lower1ex\hbox{$\sim$}}}}
\def\\{\hfil\break}

\def\lesssim{\mathrel{\hbox{\rlap{\hbox{\lower4pt\hbox{$\sim$}}}\hbox{$<$}}}}
\def\gtrsim{\mathrel{\hbox{\rlap{\hbox{\lower4pt\hbox{$\sim$}}}\hbox{$>$}}}}

\newcommand{\mamo}[1]{\mbox{$#1$}}
\newcommand{\unit}[1]{\ifmmode \:\mbox{\rm #1}\else \mbox{#1}\fi}
\newcommand{\mone}{\mamo{^{-1}}}

\newcommand{\mpc}{\unit{Mpc}}

\newcommand{\hmpc}{\mamo{h\mone}\mpc}

\begin{document}

\title [Large-scale cosmic flows from {\it Cosmicflows-2}]
{Large-scale bulk flows from the  {\it Cosmicflows-2}  catalogue}
\vskip 0.5cm
\author[Watkins and Feldman ]{Richard Watkins$^{\dagger,1}$ and Hume A. Feldman$^{\star,2}$ \\
$^\dagger$Department of Physics, Willamette University, Salem, OR 97301, USA.\\
$^\star$Department of Physics and Astronomy, University of Kansas, Lawrence, KS 66045, USA.\\
emails: $^1$rwatkins@willamette.edu;\, $^2$feldman@ku.edu}

\maketitle

\begin{abstract}

The {\it Cosmicflows-2} catalogue is a compendium of peculiar velocity measurements.   While it has many objects in common with the COMPOSITE catalogue, a previously analysed collection of peculiar velocity data found to give an unexpectedly large bulk flow on large scales, the data in {\it Cosmicflows-2} have been reanalysed to ensure consistency between distances measured using different methods.   In particular, a focus on accurate distances led the authors of the {\it Cosmicflows-2} to not correct for homogeneous or inhomogeneous Malmquist bias, both or which are corrected for in the COMPOSITE compilation.      We find remarkable agreement between the COMPOSITE and the {\it Cosmicflows-2} if the small EFAR sample of clusters located in two dense superclusters is removed from both surveys, giving results that are inconsistent with the $\Lambda$ cold dark matter standard model with Planck central parameters at the 98\% level.   On smaller scales we find overall agreement between data sets and consistency with the standard model.
\end{abstract}


\section{Introduction}
\label{sec:intro}


The large-scale bulk flow is an important cosmological probe of large scale structure.   Being the average of the peculiar velocities of the objects in a large volume, it is in principle only dependent on motions on scales which are still in the linear regime and thus can be directly related to the power spectrum of matter perturbations using linear theory.    However, in practice, the interpretation of bulk flow measurements is complicated by the large uncertainties inherent in peculiar velocity measurements, the difficulties of understanding and correcting for biases,  and by the nonuniform distributions of sample objects in peculiar velocity surveys.    Understanding and accounting for these complexities is crucial in order for the bulk flow to be useful as a cosmological probe.  

One of the main challenges in interpreting bulk flows estimated by using large-scale peculiar velocity surveys is that it is often unclear which velocity field scales they are probing.   Even the best estimations of the bulk flow were shown \citep{Kai88,WatFel95,CouWilStr00,pairwise00,NusdaCBraBer01,Hud03,pairwise03,SarFelWat07,WatFel07,FelWat08} to be affected by internal motions and other small scale effects. The scales from which the bulk flow motions arise are both large (scales larger than the sampled volume) and those of the order of, or smaller than, the volume in question. It has become clear that these effects are non-negligible and should be accounted for by the formalism \citep{optimal1,Hud03,optimal2,SarFelWat07}. While we tend to think of the bulk flow of a survey as being the motion of the volume containing the survey objects, in practice it may reflect velocities on much smaller scales. Thus, in thinking about the meaning of the bulk flow, we want to approximate the volume itself as a solid moving together, while the velocities that arise from the  forces originating inside the volume are removed by some scheme.   For example, a bulk flow calculated as the average of the velocities of a set of survey objects weighted by their peculiar velocity uncertainties has a dominant contribution from objects at smaller distances, both because they are more numerous and because their measurement uncertainties are typically much smaller.   One consequence of this confusion is that bulk flows measured using different surveys or different methods are generally not comparable, a fact that greatly complicates their use and utility as probes of large-scale structure.   

In order to standardize the study of bulk flows, we introduced the minimum variance (MV) method for obtaining estimates of bulk flows on specific scales that are comparable between surveys \citep*[][hereafter WFH]{WatFelHud09}; \citep*[][hereafter FWH]{FelWatHud10}.    The MV method of calculating the bulk flow utilizes the velocity information in a survey to estimate the flow  of an idealized, uniformly distributed set of objects with properties that can be set independent of the properties of the survey.  

The bulk flow components are in general weighted averages of measured radial peculiar velocities from a survey.   The estimated bulk flow components can be seen as convolutions of the power spectrum of the cosmic mass fluctuations with  window functions, which depend on both the spatial distribution of the sample objects as well as the weights.  The window functions thus define the scales the survey probes and tell us which scale motions contribute to the bulk flow.  In the MV method,  weights are chosen so that the resulting bulk flow estimates are as close as possible to what we would have calculated for an idealized survey.   Thus as long as the actual survey has reasonable coverage of the volume of the idealized survey, the bulk flow estimate will probe the power spectrum in a standard way, making MV bulk flow estimates comparable between different surveys.  Specifically, the MV convolution prevent small scale power leakage (aliasing) by having very little contribution from small scales where nonlinear contributions become significant, resulting in unbiased large-scale linear information.  When the window functions pick up small scale noise that leaks into the power spectrum convolution it masquerades as large scale signal. 

WFH used the MV method to analyse a compendium of available peculiar velocity data which were dubbed the COMPOSITE catalogue.      This analysis found bulk flows on scales of $100h^{-1}$ Mpc that were incompatible with the standard cosmological model at the 98-99\%  confidence level.    Some subsequent analyses using different catalogues and/or different analysis methods have agreed with the WFH results \citep{MaGorFel11,MacFelFer11,MacFelFerJaf12}, while others failed to confirm the existence of these flows; however, given the difficulties of comparing bulk flows neither have they definitely ruled them out.   In particular, \citet{MaScott13,DavNusMas11,NusBraDav11} and \citet{NusDav11} used analysis methods where it was not clear that they were probing as large a scale as WFH.   In the case of \cite{TurHudFel12}, the small size of the supernova sample they used was such that even though the MV method was used, the results were consistent with both the WFH results and the standard cosmological model.   \cite{HonSprStaScr14} analysed data from the 2MASS Tully--Fisher (TF) survey using the MV among other methods and did not find unexpectedly large flows; however, their sample is somewhat shallower than the COMPOSITE survey and they examined flows at somewhat smaller scales.   Thus the existence of large flows on the scale of $100h^{-1}$ Mpc is still an open question.   

In this paper we apply the MV formalism to the recently released {\it Cosmicflows-2} catalogue \citep{TulCouDol13}, hereafter CF2.  This catalogue contains most of the data used in the COMPOSITE catalogue of WFH, along with a significant amount of more recent data.  
The catalogue contains distances computed using several methods including the Tully--Fisher relation (TFR), Type 1a supernova (SNIa), surface brightness fluctuations, Fundamental Plane (FP), Cepheid variables, and tip of the red giant branch.   The large number of objects in the catalogue that have distances measured with more than one method makes it possible to calibrate the methods for maximum consistency.   The CF2 catalogue is notable for its size and its depth.   The group catalogue has 5,223 objects, with good coverage beyond $cz > 10,000$ km s$^{-1}$, making it well suited for measuring the bulk flow on scales of $100h^{-1}$ Mpc.    

In Section~\ref{sec:surveys} we describe the peculiar velocity samples we analyse. In Section~\ref{sec:theory} we review the formalism we utilize for the analysis. We discuss our results in Section~\ref{sec:results} and conclude in Section~\ref{sec:discussion}.
\vspace{-0.5cm}
\section{Data}
\label{sec:surveys}

The CF2 catalogue \citep{TulCouDol13} is a compendium of distances and peculiar velocities of over 8,000 galaxies, some from the literature and some from new measurements.   The majority of the galaxy distances are determined via the TFR and the FP relation, both of which give uncertainties of around 20\% of the distance, with a smaller portion of the distances coming from more accurate distance measures including SNIa, surface brightness fluctuation, Cepheids, and tip of the red giant branch.   The catalogue extends out to redshifts of 30,000 km s$^{-1}$, although it has densest coverage for the volume within 3,000 km s$^{-1}$.   
   
The COMPOSITE sample consists of various peculiar velocity catalogues. The SFI++ peculiar velocity survey of spirals in the field and in groups \citep{sfi1,sfi2,sfierr09} which consists of 2720 TF galaxies and 736 groups to make 3456 data points with characteristic depth of 35 \hmpc\footnote{Each individual survey has a characteristic MLE depth, defined as  $\sum r_q w_q / \sum w_q$ where the MLE weights are $w_q = (\sigma_q^2 + \sigma_*^2)^{-1}$. See Section~\ref{sec:theory} for more details.}.  The surface brightness fluctuation survey of \cite{TonDreBla01} with 69 field and 23 groups, with a characteristic depth of  17 \hmpc. The ENEAR survey of FP distances to nearby early-type galaxies \citep{daCBerAlo00, BerAlodaC02b, WegBerWil03} with characteristic depth of the sample is 29 \hmpc. Also included in the compilation are 103 SNeIa distances from the compilation of \cite{TonSchBar03}, limited to a distance of 150 \hmpc. The SC \citep{GioHaySal98, DalGioHay99} is a TF-based survey of spiral galaxies in 70 clusters within 200 \hmpc. The characteristic depth of the combined sample is 57 \hmpc. The SMAC sample \citep{HudSmiLuc99, HudSmiLuc04} is an all-sky  FP survey of 56 clusters. The characteristic depth of the survey is 65 \hmpc. The EFAR \citep{ColSagBur01} is a survey of 85 clusters and groups based on the FP distance indicator with a characteristic depth of 93 \hmpc. We include only the 50 clusters identified by the authors as having the best determined peculiar velocities.  \cite{Wil99b} is a TF based survey of 15 clusters with a characteristic depth of 111 \hmpc.  The COMPOSITE sample is described in detail in FWH.

The CF2 catalogue has a large number of groups and galaxies in common with the COMPOSITE compilation. An important feature of the CF2 catalogue is that the authors have utilized the large number of objects with multiple distance measurements using different methods to apply corrections to ensure consistency between data from difference sources.   While in most cases these corrections were simply shifts in zero--points, for some sets of objects more complicated adjustments were made.   

\citet{TulCouDol13} stated goal was to provide unbiased \textit{distances} at specified redshifts.   Thus they do not apply corrections for the homogeneous and inhomogeneous Malmquist bias to their sample.  Since the distances, and hence velocities,  in the COMPOSITE survey are corrected for homogenous and inhomogeneous Malmquist bias, with the one exception of the ENEAR survey, comparison of the CF2 and COMPOSITE catalogues can potentially give insight into the effect of Malmquist bias corrections on estimates of the bulk flow.   The fact that the ENEAR survey is not corrected for inhomogeneous Malmquist bias should not have a significant effect on the results of the COMPOSITE survey.   First, since the survey is shallow, any corrections would be relatively small.   Second, the large bulk flow observed in the COMPOSITE survey arises from distances $\gtrsim 70$\hmpc, where there are few ENEAR objects.  

The fact that \citeauthor{TulCouDol13} have adjusted published values for distances and peculiar velocities suggests that even though the COMPOSITE and  CF2 catalogues have many of their objects in common, what they tell us about large scale flows could potentially be very different.  However, since the catalogues do have a different spatial distribution, particularly on small scales where CF2 has many more objects, direct comparisons of the bulk flows in these catalogues are best done with methods such as the MV formalism which estimates flows that are independent of the distribution of objects. 

The CF2 data for this analysis were taken from the Extragalactic Distance Data base (EDD)\footnote{http://edd.ifa.hawaii.edu}, which at present gives distances to 8,162  individual galaxies.   For each galaxy in the data base, a distance is also given for the group, if any, that the galaxy belongs to, making it possible to create both group and galaxy versions of the CF2 catalogue.  For convenience, published distances are also given in the data base for CF2 galaxies that previously appeared in other catalogues.   

\vspace{-0.5cm}
\section{Theory}
\label{sec:theory}

Intuitively, we think of the bulk flow of a sample of objects as being the motion of the volume containing the sample.   However, in practice measured bulk flows can be much more difficult to interpret.    For example, in the maximum likelihood estimation (MLE) method \citep[e.g.][]{Kai88,Kai91,SarFelWat07,WatFel07,FelWat08,FelWat08b}, the components of the bulk flow $u_i$ are calculated as a weighted average over the measured radial velocities $S_q$ of the objects at positions $\mathbf r_q$ in the survey
\begin{equation}
u_i = \sum_{q} w_{i,q} S_q
\end{equation}
where the MLE weights, $w_{i,q}$ are given by
\begin{equation}
w_{i,q}= \sum_j A_{ij} \frac{\hat r_{q,j}}{\sigma_q^2 + \sigma_*^2}\,,
\end{equation} 
$\hat r_{q}$ is a unit vector in the direction of the $q{\rm th}$ object and
\begin{equation}
A_{ij} = \frac{\hat r_{q,i}\hat r_{q,j}}{\sigma_q^2 + \sigma_*^2}\,.
\end{equation}
In these expressions, $\sigma_q^2$ is the uncertainty in the measured peculiar velocity $S_q$, and we have introduced $\sigma_*$ to account for motion on scales much smaller than the survey.    Here the weights $w_{i,q}$ serve two purposes.   First, they account for the direction of the radial velocity relative to the bulk flow component being calculated.   Secondly, objects with larger uncertainties are down-weighted in the calculation of the bulk flow.   Since peculiar velocity uncertainties typically increase linearly with distance, objects at smaller distances generally make a larger contribution to the bulk flow.   A consequence is that the MLE bulk flow of a sample is typically more reflective of the motion of its members at smaller distances than those at the outer edge of the sample. This is a consequence of the fact that the MLE method simply minimizes the uncertainty in the flow estimate without taking into account the radial distribution of the survey objects.     

The contributions that different scales make to the bulk flow components can be quantified using linear theory.   In particular, the covariance matrix of the bulk flow components can be written as
\begin{equation}
R_{ij} \equiv \langle u_iu_j\rangle= R^{(\epsilon)}_{ij} + R^{(v)}{ij}\,,
\end{equation}
where the first term is due to noise and is given by
\begin{equation}
R_{ij}^{(\epsilon)} = \sum_q w_{i,q}w_{j,q} (\sigma_q^2 + \sigma_*^2)\,.
\end{equation}
The second term quantifies contributions from the velocity field and can be written in terms of an integral over the power spectrum $P(k)$
\begin{equation}
R_{ij}^{(v)} = \frac{\Omega_{\rm m}^{1.1}}{2{\rm \pi}^2}\int {\rm d}k\ P(k){\cal W}^2_{ij}(k)\,,
\end{equation} 
where $\Omega_m$ is the total matter fraction of the critical density and $ {\cal W}^2_{ij}(k)$ are the window functions.   Thus the diagonal window functions $ {\cal W}^2_{ii}(k)$ quantifies how sensitive the component $u_i$ is to motion on scales corresponding to the wavenumber $k$.   

The MLE method of calculating the bulk flow suffers from two major shortcomings.   First, each survey probes the power spectrum in a unique way (that is, the surveys' window functions are different) so that bulk flows calculated from different surveys are not comparable.   Secondly, the MLE window function for a given survey is fixed; it is not possible to examine different scales using the information in a single peculiar velocity catalogue.   The MV method of estimating bulk flows successfully addresses both  these deficiencies.   Unlike the MLE, the MV method uses peculiar velocity information in a sample to estimate the bulk flow of an idealized survey of galaxies whose depth and distribution are adjustable.   

To optimize the weights for the surveys, we begin by considering a hypothetical idealized survey whose moment components  $U_i$ have the desired window function that probes a desired scale.   In practice we use an ideal survey that consists of a very large number of objects isotropically distributed with a Gaussian falloff in density,  $n(r) \propto \exp(-r^2/2R_G^2)$, where $R_G$ is the scale over which the flow is analysed.  Now, suppose that we have a galaxy or cluster survey consisting of positions {\bf r}$_q$ and measured line-of-sight velocities $S_q$ with associated measurement errors $\sigma_q$.   We can calculate the weights $w_{i,q}$ that specify the moments $u_i = \sum_q w_{i,q}S_q$ that minimize the average variance, $\langle (u_i-U_i)^2\rangle$.    We call these the MV weights.  The MV moments $u_i$ calculated from these weights are the best estimates of the moments of the ideal survey, if it were to exist, that can be obtained from the available data.     We also expect that, within limits that will be described more fully below, the window functions of $u_i$ will match those of the ideal survey.  

The MLE formalism averages peculiar velocities using weights calculated to minimize the error of the flow moments. 
It does that by ignoring other essential features of the data set. In particular, it does not take into account the radial distribution of the galaxies.   The window functions of the resulting bulk flow moments will thus reflect the scales of maximum information, which will vary from survey to survey.
The MV formalism; however,  calculates weights by minimizing the theoretical variance between the estimate of the bulk flow from the actual survey and that of an ideal survey that is very dense, covers the whole sky, and has a Gaussian fall off of a particular and adjustable depth. Thus the MV scheme provides a way to find velocity moments as a function of a controllable scale ($R_G$). Further, because the MV bulk flow estimate is of a standardized quantity independent of the survey characteristics, it can be compared between different surveys.  In contrast, since the MLE formalism samples the power spectrum in a different way for each survey, direct comparison between catalogues is not possible.

As an illustration of how the MV method works, consider analyzing a typical peculiar velocity catalogue with measurement uncertainties that increase linearly with distance.    A typical survey also tends to be more dense at small distances, where measurements are easier to make.   The maximum likelihood method applied to this survey will give higher weight to nearby galaxies, where there is more information, so that the bulk flow will reflect scales somewhat smaller than that of the survey.   If one applied the MV method to the same survey, the parameter $R_G$ could be varied to examine how the bulk flow changes with depth.   As $R_G$ is increased, more weight is put on the more distant galaxies in the bulk flow estimate.    In principle, the downside of changing the weights is that it increases the uncertainty in the resulting estimate; however, we have found that in reasonable applications of the MV method this increase in uncertainty has not been significant.    Specifically, the MV method works well when the volume of the idealized survey is well populated with objects from the peculiar velocity sample.   

\begin{figure}
  \begin{center}
\includegraphics[scale=0.4]{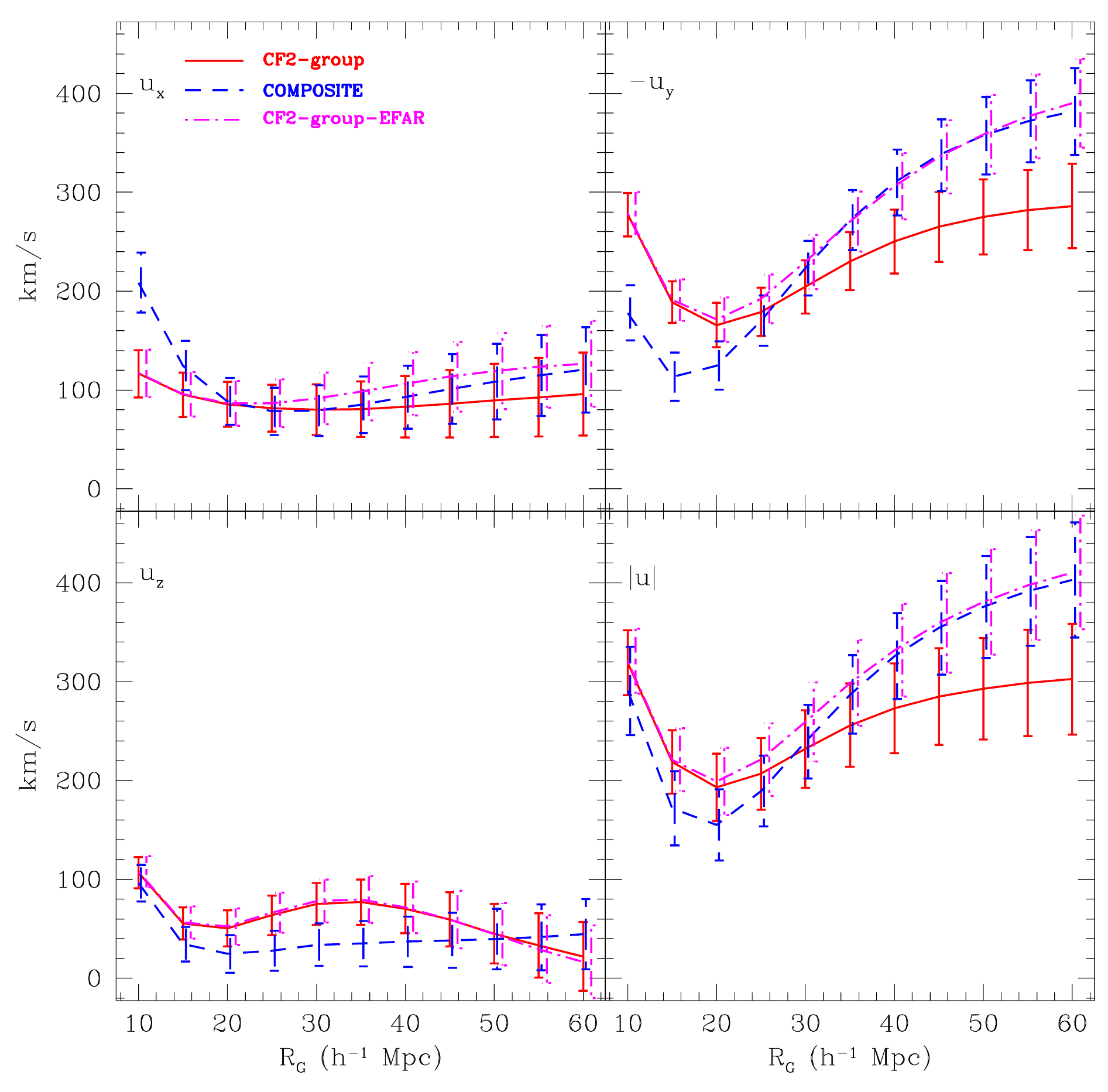}
 \end{center}
 \vspace{-0.5cm}
\caption{\small
The estimates of the MV BF of the CF2 (red solid lines) and COMPOSITE (blue dashed) catalogues as a function of $R_G$ in galactic coordinates.
We also show the results for the CF2--EFAR compilation (magenta dash--dotted) which agrees remarkably well with the COMPOSITE results.
}
\label{fig:bf}
\end{figure}

\vspace{-0.5cm}
\section{Results}
\label{sec:results}

\begin{table*}
\caption{Bulk flow vectors for the surveys (in Galactic Cartesian coordinates) for MV weights for $R_G=50$\hmpc.  The quoted errors includes both noise and the difference from the idealized survey geometry. The last two columns are total observed probability $P(>\chi^2)$ of finding flows as large or larger, in percentage, for the central parameters $(h,\Omega_m,\sigma_8)$ from WMAP9 \citep[$(0.700,  0.279,   0.821)$]{WMAP9} and Planck \citep[$(0.671  , 0.318 ,  0.834)$]{Planckparameters13}. For the top list, we used the estimated distances for the positions, the bottom list we used the redshifts rather than estimated distances to specify the positions of the objects in the catalogues.} 
\begin{tabular}{lcccccrr}
 \hline \hline
Survey      & $N$  & $u_x$ & $u_y$ & $u_z$  & $|u|$ &  $P_{\rm WMAP}$ &  $P_{\rm Planck}$   \\
& & (km s$^{-1}$) & (km s$^{-1}$) & (km s$^{-1}$) & (km s$^{-1}$) & (\%)\,\,\,\,\,\, &(\%)\,\,\,\,\,\, \\ \hline \hline \\
\multicolumn{8}{c}{\bf Estimated Distances}\\ \\
${\rm COMPOSITE}$ &  4530 &      101.0 $\pm$     38.3 &     -362.1 $\pm$     39.2 &       39.3 $\pm$     30.6 &      377.9 $\pm$     62.7 &   1.9 &   1.6 \\ 
CF2$_{\rm group}$ &  4845 &       88.4 $\pm$     37.0 &     -274.6 $\pm$     37.9 &       45.3 $\pm$     29.9 &      292.0 $\pm$     60.8 &  10.5 &   9.4 \\ 
CF2$_{\rm galaxy}$ &  8094 &      115.6 $\pm$     36.5 &     -261.1 $\pm$     37.4 &       43.3 $\pm$     29.4 &      288.8 $\pm$     59.9 &  18.4 &  18.5 \\ \\
${\rm COMPOSITE-EFAR}$ &  4485 &       86.8 $\pm$     38.7 &     -383.0 $\pm$     39.7 &       29.2 $\pm$     31.2 &      393.8 $\pm$     63.7 &   1.4 &   1.2 \\ 
CF2$_{\rm group}-EFAR$ &  4727 &      118.1 $\pm$     38.5 &     -358.2 $\pm$     39.9 &       45.0 $\pm$     31.7 &      379.9 $\pm$     63.9 &   1.8 &   1.5 \\ 
CF2$_{\rm galaxy}-EFAR$ &  7431 &      126.0 $\pm$     37.3 &     -329.9 $\pm$     38.5 &       31.4 $\pm$     30.3 &      354.5 $\pm$     61.6 &   3.0 &   2.6 \\ \\
CF2$^{\rm adjusted}_{\rm group}$ &  4845 &       80.6 $\pm$     37.0 &     -244.3 $\pm$     37.9 &       41.9 $\pm$     29.9 &      260.7 $\pm$     60.8 &  17.9 &  16.5 \\ 
CF2$^{\rm adjusted}_{\rm galaxy}$ &  8094 &      103.5 $\pm$     36.5 &     -229.5 $\pm$     37.4 &       41.8 $\pm$     29.4 &      255.2 $\pm$     59.9 &  22.2 &  22.2 \\ \\
CF2$^{\rm adjusted}_{\rm group}-EFAR$ &  4727 &      108.1 $\pm$     38.5 &     -313.7 $\pm$     39.9 &       43.2 $\pm$     31.7 &      334.7 $\pm$     63.9 &   5.0 &   4.3 \\ 
CF2$^{\rm adjusted}_{\rm galaxy}-EFAR$ &  7431 &      113.3 $\pm$     37.3 &     -290.4 $\pm$     38.5 &       31.5 $\pm$     30.3 &      313.3 $\pm$     61.6 &   7.3 &   6.4 \\ \\
EFAR &    50 &      548.9 $\pm$    346.7 &      489.0 $\pm$    262.1 &       10.3 $\pm$    225.9 &      735.3 $\pm$    489.8 &  15.5 &  15.7 \\ 
EFAR$_{\rm CF2}$ &   113 &      168.1 $\pm$    310.7 &      109.2 $\pm$    209.7 &      161.8 $\pm$    206.1 &      257.6 $\pm$    427.7 &  56.3 &  56.2 \\ 
\hline \\
\multicolumn{8}{c}{\bf Redshifts}\\ \\
${\rm COMPOSITE}$ &  4530 &       25.7 $\pm$     36.9 &     -367.7 $\pm$     37.8 &       38.2 $\pm$     29.8 &      370.6 $\pm$     60.6 &   2.1 &   2.1 \\ 
CF2$_{\rm group}$ &  4845 &       50.5 $\pm$     36.4 &     -245.8 $\pm$     37.5 &       39.7 $\pm$     29.9 &      254.0 $\pm$     60.2 &  19.7 &  18.1 \\ 
CF2$_{\rm galaxy}$ &  8093 &       28.6 $\pm$     35.7 &     -276.7 $\pm$     36.9 &       74.0 $\pm$     29.1 &      287.9 $\pm$     59.0 &  10.7 &   9.6 \\ \\
${\rm COMPOSITE-EFAR}$ &  4480 &        9.8 $\pm$     37.4 &     -368.0 $\pm$     38.3 &       26.8 $\pm$     30.4 &      369.2 $\pm$     61.6 &   2.3 &   1.9 \\ 
CF2$_{\rm group}-EFAR$ &  4727 &       74.9 $\pm$     37.7 &     -279.0 $\pm$     39.5 &       37.5 $\pm$     31.6 &      291.3 $\pm$     63.1 &  11.5 &  10.4 \\ 
CF2$_{\rm galaxy}-EFAR$ &  7431 &       33.4 $\pm$     36.4 &     -334.8 $\pm$     38.1 &       60.3 $\pm$     30.0 &      341.8 $\pm$     60.7 &   3.9 &   3.4 \\ \\
CF2$^{\rm adjusted}_{\rm group}$ &  4845 &       49.0 $\pm$     36.4 &     -226.5 $\pm$     37.5 &       35.6 $\pm$     29.9 &      234.4 $\pm$     60.2 &  26.3 &  24.6 \\ 
CF2$^{\rm adjusted}_{\rm galaxy}$ &  8093 &       28.7 $\pm$     35.7 &     -255.5 $\pm$     36.9 &       70.2 $\pm$     29.1 &      266.6 $\pm$     59.0 &  15.6 &  14.2 \\ \\
CF2$^{\rm adjusted}_{\rm group}-EFAR$ &  4727 &       72.9 $\pm$     37.7 &     -251.5 $\pm$     39.5 &       36.5 $\pm$     31.6 &      264.4 $\pm$     63.1 &  18.0 &  16.6 \\ 
CF2$^{\rm adjusted}_{\rm galaxy}-EFAR$ &  7431 &       34.0 $\pm$     36.4 &     -307.8 $\pm$     38.1 &       59.0 $\pm$     30.0 &      315.2 $\pm$     60.7 &   6.8 &   6.0 \\ \\
EFAR &    50 &      502.8 $\pm$    346.6 &      527.7 $\pm$    262.0 &       52.2 $\pm$    225.6 &      730.8 $\pm$    489.6 &  14.4 &  14.7 \\ 
EFAR$_{\rm CF2}$ &   113 &       71.0 $\pm$    310.3 &      192.8 $\pm$    209.5 &      235.2 $\pm$    205.6 &      312.3 $\pm$    427.2 &  44.0 &  44.0 \\ 
\label{tab:bf}
\end{tabular}
\end{table*}

In Table~\ref{tab:bf} we show the estimates of the MV bulk flow components for $R_G=50h^{-1}$ Mpc for the COMPOSITE sample and for grouped and ungrouped versions of the CF2 compilation (see also Fig.~\ref{fig:bf}).   Here we see that the magnitude of the bulk flow is significantly smaller in the CF2 catalogues.   In order to better understand the difference between the bulk flow calculated using the  CF2 and the COMPOSITE we broke each survey into its component surveys and did detailed comparisons between them.   As in WFH, we found good agreement between the bulk flows of the component surveys, with the EFAR survey being the one notable exception.   As shown in Table~\ref{tab:bf}, the MV bulk flow components of the EFAR survey differ greatly from the other surveys.   In particular, while the bulk flow of all other surveys has a significant negative $y$-component, the EFAR survey bulk flow has a  positive $y$-component (with very large errors).   It is not too surprising that the EFAR survey should give different results given the fact that, unlike the other component surveys, it is not full sky, but rather was designed to study motions in the two superclusters Hercules-Corona Borealis and Perseus-Pisces-Cetus.   Since EFAR consists of a relatively small number of clusters drawn from highly atypical, anisotropically distributed regions, it seems unlikely that these objects would accurately reflect the large-scale flow.   In addition,  the large bulk flow seen in the COMPOSITE survey is mostly due to a large negative $y$-component and the EFAR clusters are particularly unbalanced in that direction, having some clusters near the $+y$ direction and none in the $-y$ direction.  A related issue is one of selection bias.   Our goal is to sample the volume in which we are calculating the bulk flow as isotropically as possible.   Including a highly anisotropic set of clusters will result in some directions having more weight than others in our analysis, resulting in a potentially biased bulk flow.

The effect of the inclusion of the EFAR survey on the bulk flow can be seen in Table~\ref{tab:bf}, where we also show the bulk flow for the CF2 and COMPOSITE compilations  with EFAR removed (CF2$_{\rm group}$--EFAR and COMPOSITE--EFAR, respectively).   We see that once the EFAR survey is taken out, we have remarkable agreement between the two surveys (this effect can also be clearly seen in Fig.~\ref{fig:bf}).   One question that arises is why the EFAR survey has a substantially larger effect on the CF2 results than on the COMPOSITE?    There are two reasons for this.  First, the CF2 includes the entire set of 113 clusters included in Table 2 of \citet{ColSagBur01}, including foreground and background galaxies, while the COMPOSITE includes only the 50 best measured clusters used in the paper for peculiar velocity analysis.   Second, \citet{TulCouDol13} includes additional distances for galaxies in the EFAR clusters taken with other distance indicators.   This, together with differences in how uncertainties are calculated, results in the EFAR cluster peculiar velocities having significantly smaller uncertainties in the CF2 than in the COMPOSITE, giving them correspondingly greater weight in the analysis.  

Table~\ref{tab:bf} shows that the disagreement between the CF2 and the COMPOSITE survey is primarily due to the EFAR clusters and the greater weight that they carry in the CF2 survey.   We also show the bulk flow calculated for the CF2 galaxy catalogue with and without the EFAR sample.   Note that we have only removed galaxies with distances measured using the FP reported in \citep{ColSagBur01}; distance measurements made using other methods for galaxies in EFAR clusters have been retained.   This explains the slightly smaller bulk flow measured in the CF2$_{\rm galaxy}$--EFAR relative to the CF2$_{\rm group}$--EFAR.   In the following we will focus on the COMPOSITE--EFAR and CF2$_{\rm group}$--EFAR samples.   

\begin{figure}
  \begin{center}
\includegraphics[scale=0.41]{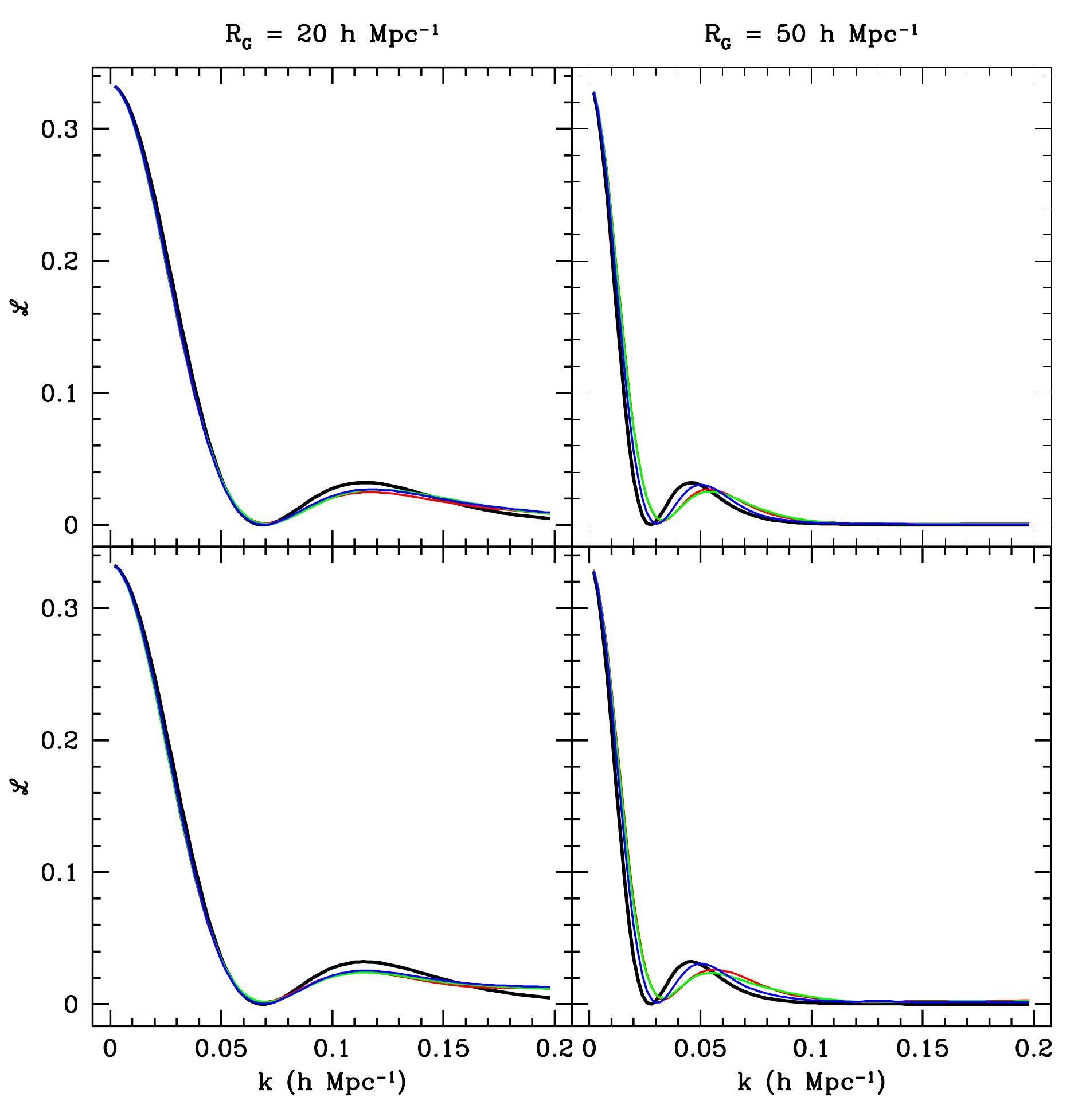}
 \end{center}
 \vspace{-0.5cm}
\caption{\small
The normalized window functions of the bulk flow component for $R_G=20$\hmpc\ (left--hand panels) and $R_G=50$\hmpc\ (right--hand panels) for the CF2$_{\rm group}$--EFAR catalogue (upper panels) and the COMPOSITE--EFAR (lower panels). The thin lines are the window functions for the MV for each of the bulk flow Galactic Cartesian components ($x$ - red, $y$ - green, $z$ - blue) respectively. The thick black line is the ideal window function (since the ideal survey is isotropic, all component are the same). It is clear that the window functions for the two samples agree with each other very well.
}
\label{fig:window}
\end{figure}

In Fig.~\ref{fig:window} we show the window functions for the MV estimates of the three components of the bulk flow for the two surveys for $R_G=20$ (left--hand panels) and $50h^{-1}$ Mpc (right--hand panels), together with the window function for the ideal survey for the CF2$_{\rm group}$--EFAR catalogue (upper panels) and the COMPOSITE--EFAR (lower panels).   The window functions agree with each other quite well, suggesting that the MV estimates from the two surveys probe the power spectrum in a similar way and that both surveys' window functions are good matches to the idealized survey. In Fig.~\ref{fig:bf} we show the MV bulk flow estimates for both the CF2$_{\rm group}$--EFAR and the COMPOSITE--EFAR surveys as a function of the scale $R_G$.   Again, we see remarkable agreement between the two surveys, once the EFAR points are taken out. 
\begin{figure}
  \begin{center}
\includegraphics[scale=0.41]{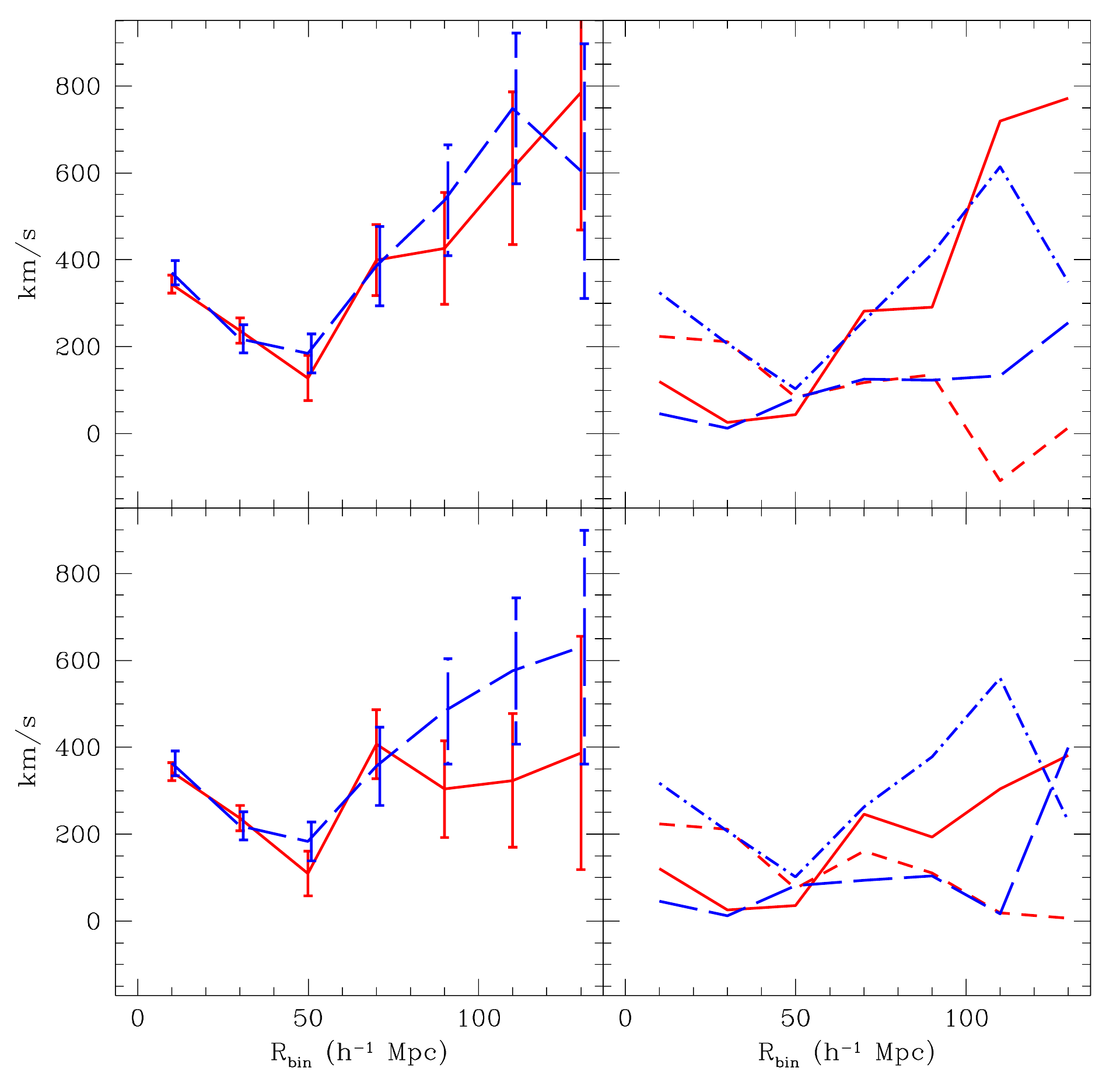}
 \end{center}
 \vspace{-0.5cm}
\caption{\small
In the top left--hand panel we show the maximum likelihood bulk flow component $-u_y$ for $20h^{-1}$Mpc thick shells for the CF2$_{\rm group}$--EFAR (red solid line) and the COMPOSITE--EFAR (blue long-dash) surveys. In the top right--hand panel we show the same information as in the left--hand panel except that we have given the contribution to the bulk flow in the (CF2$_{\rm group}$--EFAR, COMPOSITE--EFAR) surveys from galaxies with $y>0$ (red solid, blue long-dash, respectively) and $y<0$ (red short-dashed, blue dash-dot, respectively) separately. The bottom panels show the same as the top panels with the EFAR clusters included in the compilations.
}
\label{fig:shells}
\end{figure}

In order to investigate the origin of the larger than expected bulk flow in our two samples, we examined the radial dependence of the bulk flows in more detail.   In Fig. ~\ref{fig:shells} (top left--hand panel) we show the maximum likelihood bulk flow component $-u_y$ for $20h^{-1}$Mpc thick shells for both surveys.   Maximum likelihood estimates work well for shells since the objects in a shell are all at similar distances, so there is no issue with having a radius dependent weighting as there is in the case of spherical volumes.      In the figure we see that in both surveys, the shell $-u_y$ initially drops with radius, but then at about  50 $h^{-1}$Mpc, $-u_y$ turns around and begins to increase.   

Given that we are working with the radial components of peculiar velocities, contributions to the bulk flow component $u_y$ primarily come from objects near the $\pm y$ directions.   It is interesting to separate the contributions to the bulk flow coming from these two regions.   In the top right--hand panel of Fig.~\ref{fig:shells}
we show the same information as in the left--hand panel except that we have shown the contribution to the bulk flow from galaxies with $y>0$ and $y<0$ separately.   For both surveys, we see that the anomalously large $-u_y$ has negative contributions from both sides of the sky, so that at $y>0$ galaxies are streaming towards us, and hence have a negative radial peculiar velocity, while at $y<0$ galaxies are moving away, and hence have a positive radial peculiar velocity.    However, it is clear that the dominant contribution to $u_y$ comes from positive peculiar velocities of objects in  the $y<0$ direction.  Similarly, the contribution from the $y>0$ direction is due to a somewhat smaller excess of negative peculiar velocities.

\begin{figure}
  \begin{center}
\includegraphics[scale=0.41]{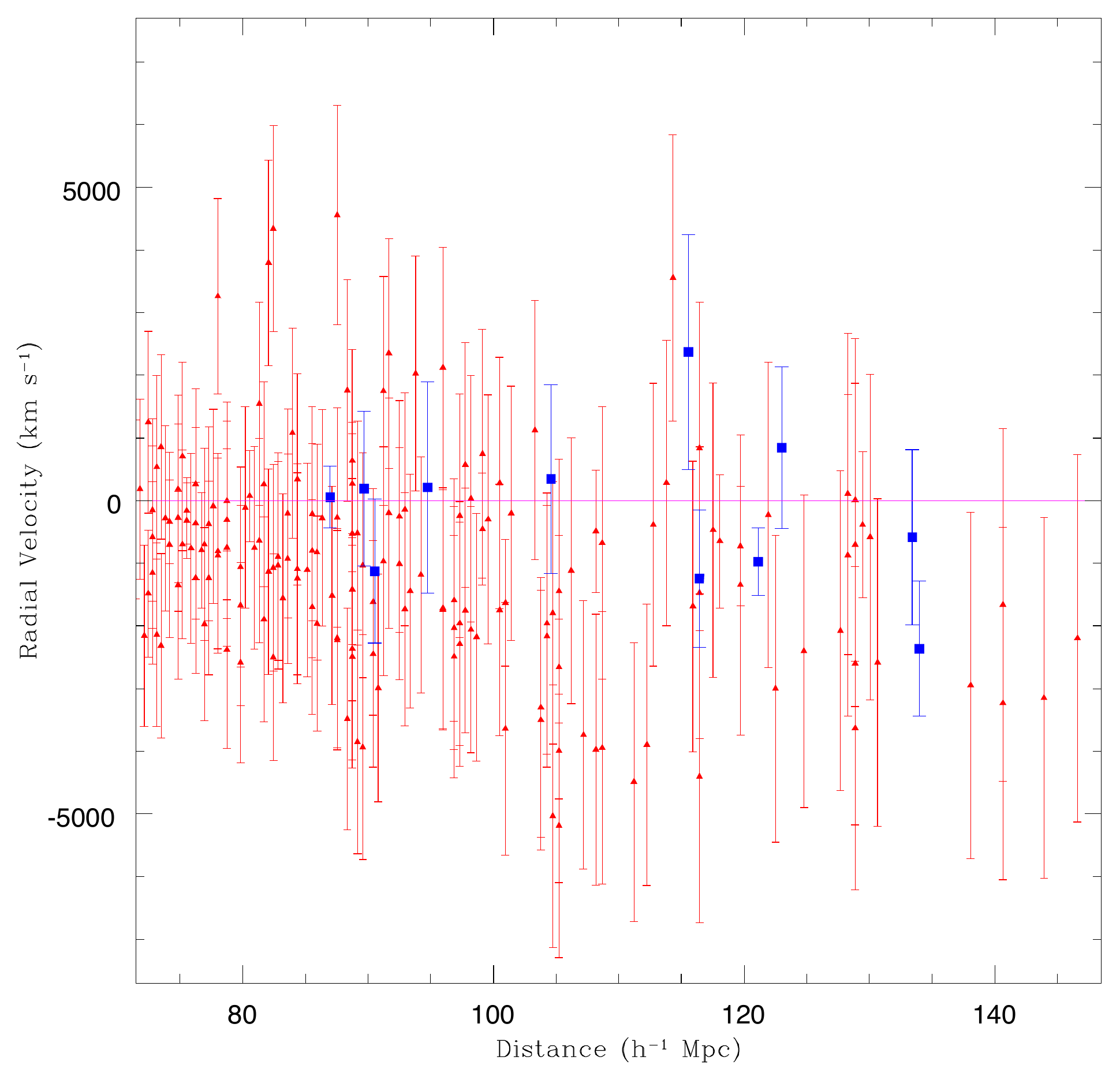}
 \end{center}
 \vspace{-0.5cm}
\caption{\small
The red dots are the radial peculiar velocities of the CF2$_{\rm group}$--EFAR data points, while the blue squares are the CF2 radial peculiar velocities of the EFAR clusters as a function of distance in the $+y$ direction. As mentioned in the text, there are no EFAR clusters in the $-y$ direction.
}
\label{fig:plusy}
\end{figure}

\begin{table*}
\caption{Same as Table~\ref{tab:bf} for $R_G=20$\hmpc. Here we show the results only for the estimated distances as the redshifts for close-by objects are not a good approximation for distance.} 
\begin{tabular}{lcccccrr}
 \hline \hline
Survey      & $N$  & $u_x$ & $u_y$ & $u_z$  & $|u|$ &  $P_{\rm WMAP}$ &  $P_{\rm Planck}$   \\
& & (km s$^{-1}$) & (km s$^{-1}$) & (km s$^{-1}$) & (km s$^{-1}$) & (\%)\,\,\,\,\,\, &(\%)\,\,\,\,\,\, \\ \hline \hline \\
\multicolumn{8}{c}{\bf Estimated Distances}\\ \\
${\rm COMPOSITE}$ &  4530 &       71.9 $\pm$     23.7 &     -132.7 $\pm$     24.4 &       28.0 $\pm$     19.1 &      153.5 $\pm$     39.0 &  83.4 &  83.4 \\ 
CF2$_{\rm group}$ &  4845 &       84.2 $\pm$     22.7 &     -165.5 $\pm$     22.4 &       50.8 $\pm$     18.4 &      192.5 $\pm$     36.8 &  71.5 &  71.5 \\ 
CF2$_{\rm galaxy}$ &  8093 &       67.9 $\pm$     21.6 &     -137.5 $\pm$     21.5 &       40.5 $\pm$     17.2 &      158.6 $\pm$     35.0 &  82.0 &  82.0 \\ \\
${\rm COMPOSITE-EFAR}$ &  4485 &       69.4 $\pm$     23.7 &     -135.2 $\pm$     24.4 &       28.6 $\pm$     19.1 &      154.7 $\pm$     39.0 &  83.1 &  83.1 \\ 
CF2$_{\rm group}-EFAR$ &  4727 &       85.3 $\pm$     22.8 &     -171.1 $\pm$     22.5 &       52.6 $\pm$     18.4 &      198.3 $\pm$     36.9 &  69.6 &  69.6 \\ 
CF2$_{\rm galaxy}-EFAR$ &  7431 &       69.2 $\pm$     21.6 &     -144.1 $\pm$     21.5 &       41.5 $\pm$     17.2 &      165.1 $\pm$     35.1 &  80.2 &  80.2 \\ \\
CF2$^{\rm adjusted}_{\rm group}$ &  4845 &       77.1 $\pm$     22.7 &     -147.6 $\pm$     22.4 &       47.1 $\pm$     18.4 &      173.0 $\pm$     36.8 &  77.7 &  77.7 \\ 
CF2$^{\rm adjusted}_{\rm galaxy}$ &  8093 &       60.3 $\pm$     21.6 &     -124.2 $\pm$     21.5 &       39.8 $\pm$     17.2 &      143.7 $\pm$     35.0 &  86.0 &  86.0 \\ \\
CF2$^{\rm adjusted}_{\rm group}-EFAR$ &  4727 &       78.2 $\pm$     22.8 &     -151.9 $\pm$     22.5 &       49.1 $\pm$     18.4 &      177.8 $\pm$     36.9 &  76.3 &  76.3 \\ 
CF2$^{\rm adjusted}_{\rm galaxy}-EFAR$ &  7431 &       61.6 $\pm$     21.6 &     -130.1 $\pm$     21.5 &       40.9 $\pm$     17.2 &      149.7 $\pm$     35.1 &  84.5 &  84.5 \\ \\
EFAR &    50 &      545.0 $\pm$    347.1 &      494.5 $\pm$    262.7 &        1.1 $\pm$    226.1 &      735.9 $\pm$    490.5 &  15.9 &  15.9 \\ 
EFAR$_{\rm CF2}$ &   113 &      173.1 $\pm$    311.8 &      137.9 $\pm$    210.4 &      148.7 $\pm$    207.7 &      266.7 $\pm$    429.7 &  56.8 &  56.8 \\ 
\label{tab:bf20}
\end{tabular}
\end{table*}

In Fig.~\ref{fig:shells} (bottom panels) we also show results for the full CF2$_{\rm group}$ and COMPOSITE compilations, that is, with the EFAR clusters included.   We see that the effect of the EFAR clusters is to reduce the large negative velocities in the $+y$ direction at distances greater than about $70h^{-1}$Mpc.   In Fig.~\ref{fig:plusy} we show the CF2 radial peculiar velocities (with uncertainties) of galaxies and clusters within $45^o$ of the $+y$ direction as a function of distance.   We see that the EFAR clusters have generally more positive velocities than the other objects.    Even though there are only a few EFAR clusters, their small uncertainties in the CF2 give them substantial weight in the analysis.   The disagreement between velocities of the EFAR clusters and the other survey objects in the surrounding volume could be due to the fact that the small number of EFAR clusters sample only a small region of the volume, while the larger number of other objects sample the volume more uniformly.   It is also possible that clusters on the near side of the supercluster were preferentially selected, so that the positive velocities reflect infall into the superclusters.  

\citeauthor{TulCouDol13} find that the velocities in the CF2 are skewed towards the negative.  They attribute this to the fact that while uncertainties in distance moduli are Gaussian, uncertainties in distances will follow a skewed distribution, resulting in error-induced peculiar velocities that tend to be more negative than positive, an effect that they call \textit{error bias}.   Since peculiar velocity errors grow linearly with distance, this potential bias will have the greatest effect at large distances, where we have seen that the large bulk flow in the surveys is originating.     They suggest correcting for this bias by leaving positive velocities unchanged, but adjusting negative peculiar velocities $V_{pec}$ using the formula
\begin{equation}
V_{adj}= V_{pec}\left[0.77 + 0.23e^{-0.01(e_dV_{mod})}\right] \,,
\end{equation}
where $e_d$ is the fractional distance uncertainty and $V_{mod}$ is the modified value of $cz$ for the galaxy or cluster as discussed above.   While it is not clear that this is an effective way to correct for this bias, for comparison we have also included in Table~\ref{tab:bf} results for the CF2 catalogues when this adjustment is applied.  It makes sense that this adjustment results in a smaller bulk flow, since as discussed above, the large bulk flow is mostly due to negative velocities in the $+y$ direction.     Finally, in the bottom of Table~\ref{tab:bf} we show results for the case where redshifts are used instead of distances to specify the positions of objects in the samples. 

For all the catalogues in Table~\ref{tab:bf} we show the probability of obtaining a bulk flow as large or larger assuming the standard $\Lambda$ cold dark matter model with $WMAP$9 or Planck parameters (see table 3 in WFH and table 4 in FWH).

Finally, in Table~\ref{tab:bf20} we show the same information as in Table~\ref{tab:bf} except for $R_G= 20$\hmpc.  Here we see remarkable agreement between all the catalogues, with the exception of EFAR, which has very little information on these scales.  We also emphasize that on these small scales there is strong consistency of all the data with the expectations of the standard cosmological model with parameters taken from microwave background observations.    

\vspace{-0.5cm}
\section{Discussion}
\label{sec:discussion}

WFH and FWH reported a larger than expected bulk flow on scales of 100 \hmpc\ based on a compilation of distance and peculiar velocity data dubbed the COMPOSITE survey.   \citet{TulCouDol13} presented a new compilation of peculiar velocity data, CF2, which both adds new data and adjusts published distances and peculiar velocities to achieve consistency between the constituent data sets that use a variety of different distance measurement techniques.  In this paper we have analysed the large scale motions in the CF2 samples and found that, if the EFAR cluster sample is removed, there is remarkable agreement between the catalogues.   This suggests that the bulk flow estimate is robust with regard to the details of the analysis, and in particular, whether corrections are made for homogeneous and inhomogeneous Malmquist bias.    

Using the MV method it is possible to estimate the bulk flow on different scales using a given set of peculiar velocity measurements.   A puzzling feature of the flows reported by WFH and seen here in the CF2-EFAR compilations is that they are quite consistent with expectations on scales $\lessim 40h^{-1}$ Mpc (corresponding to $R_G\lessim 20h^{-1}$ Mpc), but then the bulk flow increase with scale as $R_G$ increases.   Indeed, several studies have shown that peculiar motions on scales $\lessim 40h^{-1}$ Mpc are consistent with expectations from observations of the density field 
\citep{PikHud05,ErdLahHuc06,BilChoJarMam11,NusBraDav11,MaBraSco12}. It is difficult to explain physically how a smaller volume can not participate in the flow of the larger volume that contains it.  

As we have seen, COMPOSITE and  CF2 bulk flows agree on small scales, but disagree on larger scales if the EFAR clusters are included in the samples.   This is due to the fact that the EFAR clusters carry substantially more weight in the analysis of the CF2 samples due to their greater number and their significantly smaller uncertainties.   We suggest that the disagreement between the peculiar velocities of the EFAR clusters and the other survey objects in the same volume is due to the fact that the EFAR clusters sample only small regions of the volume, that they do so in a highly anisotropic manner, and that these small regions are atypical due to being in the proximity of a dense supercluster.   

We have explored the effect of \textit{error bias} on the CF2 bulk flow estimate.   Error bias tends to amplify negative velocities, and since we have seen that the unexpectedly large bulk flow reported by WFH is mostly due to negative velocities in the $+y$ direction, we expect that adjusting the velocities to correct for this effect should reduce the magnitude of this bulk flow.   This is indeed the case, and we see that when the adjusted data is used the probability of obtaining a bulk flow as large or larger in the standard model becomes of the order of 5 per cent ($\sim2\sigma$).   While still small, this reduces the disagreement with theory to a level that is easier to dismiss as a statistical fluctuation.   However, we feel that more work needs to be done to show that this adjustment is an effective correction for error bias.   

There are still some open questions about the measurement and analysis of large scale peculiar velocities.  First, \cite{TulCouDol13} use the variation of the Hubble parameter with redshift as an indicator of bias.   However, this assumes that the Hubble parameter is the same in all regions of space, an assumption that has been questioned by various authors \citep{GioDalHayHar99,ConCarGuyHow07,SinDavHau10,WilSmaMatWat12}.   Furthermore, a  bulk flow can itself cause an apparent spatial variation in the Hubble constant.   By adjusting all of the subsets of the CF2 catalogue to have the same Hubble constant, \citeauthor{TulCouDol13}  may have inadvertently reduced the magnitude of a real bulk flow, resulting in the generally smaller, although consistent,  bulk flow values found in the CF2 catalogue relative to the COMPOSITE .   

A related issue is that \cite{TulCouDol13} say that their catalogue is consistent with a value for the Hubble parameter of $74.4\pm3.0$ km s$^{-1}$Mpc$^{-1}$, a value that is in conflict with the recent Planck measurement of $67.4\pm1.4 $ km s$^{-1}$Mpc$^{-1}$  \citep{Planckparameters13}, which is lower than other local measurements as well.  Since peculiar velocity determinations depend crucially on understanding the characteristics of cosmological redshift,
until the discrepancy between local and global measurements of the Hubble parameter is resolved, we cannot be confident that we truly understand motions on scales of $\gtrsim(100$\hmpc).   

\vspace{0.5cm}

\noindent{\bf ACKNOWLEDGEMENTS:} We would like to thank Mike Hudson, Adi Nusser, and Brent Tully for useful conversations.

\vspace{-0.9cm}

\bibliographystyle{mn2e}
\bibliography{haf}

\end{document}